\documentclass[aps,prl,superscriptaddress,twocolumn,10pt]{revtex4-2}

\usepackage{amssymb}
\usepackage{cancel}
\usepackage{color,graphicx}
\usepackage{amsmath,lmodern}
\usepackage{amsbsy}
\usepackage{amsthm}
\usepackage{bbm}
\usepackage{epsfig}
\usepackage{float}
\usepackage{graphicx}
\usepackage{subfigure}
\usepackage{dcolumn}
\usepackage{bbm}
\usepackage{xcolor,epstopdf}
\usepackage{amscd}
\usepackage{amsfonts}  
\usepackage[]{amsmath}
\usepackage{amssymb}    
\usepackage{mathrsfs}
\usepackage{verbatim}
\usepackage[]{cases}
\usepackage{amsmath}
\usepackage{wasysym}
\usepackage[T1]{fontenc}
\usepackage{mathtools}
\usepackage{todonotes}
\usepackage[normalem]{ulem}
\usepackage[colorlinks=true,linkcolor=blue,urlcolor=blue,citecolor=blue,pdfusetitle]{hyperref}
\usepackage{cleveref}
\usepackage{libertine}
\usepackage{hyperref}
\usepackage{comment}

\usepackage{amssymb,amsmath,bbm}
\usepackage[utf8]{inputenc}
\usepackage{comment,bm}
\usepackage{braket}
\usepackage[normalem]{ulem}
\usepackage{dcolumn}

\begin{document}

\title{Diagnostics of quantum-gate coherences via end-point-measurement statistics}

\author{Ilaria Gianani} 
\affiliation{Dipartimento di Scienze, Universit\'a degli Studi Roma Tre, Via della Vasca Navale, 84, 00146 Rome, Italy}

\author{Alessio Belenchia} 
\affiliation{Institut f\"ur Theoretische Physik, Eberhard-Karls-Universit\"at T\"ubingen, 72076 T\"ubingen, German}
\affiliation{Centre for Quantum Materials and Technology, School of Mathematics and Physics, Queen's University Belfast, Belfast BT7 1NN, United Kingdom}

\author{Stefano Gherardini} 
\affiliation{Istituto Nazionale di Ottica -- CNR, Area Science Park,Basovizza, I-34149 Trieste, Italy}
\affiliation{LENS, University of Florence, via Carrara 1, I-500}

\author{Vincenzo Berardi}
\affiliation{Dipartimento Interateneo di Fisica “Michelangelo Merlin”, Politecnico di Bari, Via Orabona 4, 70126 Bari, Italy}

\author{Marco Barbieri} 
\affiliation{Dipartimento di Scienze, Universit\'a degli Studi Roma Tre, Via della Vasca Navale, 84, 00146 Rome, Italy}
\affiliation{Istituto Nazionale di Ottica -- CNR, Largo Enrico Fermi 6, 50125 Florence, Italy}

\author{Mauro Paternostro}
\affiliation{Centre for Quantum Materials and Technology, School of Mathematics and Physics, Queen's University Belfast, Belfast BT7 1NN, United Kingdom}

\begin{abstract}
Quantum coherence is a central ingredient in quantum physics with several theoretical and technological ramifications. In this work we consider a figure of merit encoding the information on how the coherence generated on average by a quantum gate is affected by unitary errors (coherent noise sources). We provide numerical evidences that such information is well captured by the statistics of local energy measurements on the output states of the gate. These findings are then corroborated by experimental data taken in a quantum optics setting.  
\end{abstract}

\maketitle


The characterization of the {\it quality} of a given operation is a key step towards the validation of quantum technologies: any information-processing task needs to be assessed against relevant performance-quantifying figures of merit. A relevant instance of such key necessity is the evaluation of the quality of quantum gates to assess fault tolerance~\cite{Gottesman2010,Ladd2010}.

Quantum coherence is a pivotal property of quantum states and operations~\cite{Streltsov}, and embodies the ultimate feature setting quantum and classical mechanics apart. Its presence in  resource states or generation through dynamics are paramount to the achievement of {\it quantum advantages}~\cite{BaumgratzPRL2014}. Evidences in this respect have been provided for transport across biomolecular networks~\cite{bio1,bio2,bio3,bio4,bio5,bio6}, nano-physics~\cite{nano1,nano2} and low-temperature thermodynamics \cite{thermo1,thermo2,thermo3,thermo4,thermo5,thermo6}. Recently formulated resource-theory
approaches~\cite{Streltsov,BaumgratzPRL2014,Yadin,Dana} have enabled the assessment of the role of quantum coherence in a range of tasks in quantum technologies~\cite{Hillery2016,Matera2016,Giorda2018,Streltsov2016}. As quantum coherence is closely related to and can be converted into other powerful quantum resources~\cite{Ma2016,Wu2017,Wu2018}, its characterization is growing both in interest and relevance. In Ref.~\cite{Yuan2020}, a scheme for the direct estimation of quantum coherence through the implementation of entangled measurements on two copies of the state at hand has been experimentally demonstrated in a linear-optics platform. While providing an interesting approach to the quantification of coherences without relying on a tomographic approach based on the reconstruction of the system state and the {\it evaluation} of a suitable quantitative measure~\cite{Streltsov,BaumgratzPRL2014}, this approach poses challenges embodied by the need for entangled measurements to be used. 

In this paper we combine the need for the characterization of quantum coherence in quantum processes and states with the observation that any information-processing task implies a process of energy exchange between the elements of the computational register. In this regard, we make use of tools specifically designed to address the phenomenology of energetics in non-equilibrium quantum processes to build a diagnostic instrument for the quality of quantum coherence resulting from a quantum gate. We unveil a remarkable connection between the statistics of local observables -- focusing in particular on energy measurements~\cite{CimininpjQI2020} -- and the coherence induced by the action of a quantum gate, in the cornerstone example of control-unitary two-qubit gates. In order to provide a figure of merit that accounts for the coherence-inducing capabilities of the gate at hand irrespectively of the specific input states, we focus on an {\it input-averaged} quantity that has no need for the tomographic reconstruction of the computational register states. Our results, albeit surprising, are well aligned with a stream of recent works in which non-equilibrium thermodynamic relations have been used as a diagnostic tool for the non-unitarity and dissipation of commercial quantum computing architectures~\cite{GardasSciRep2018,BuffoniQST2020,CimininpjQI2020,CampisiPRE2021,StevensArXiv2021,Fellous-AsianiPRXQuantum2021,BuffoniArXiv2022}. 


\noindent
\textbf{\textit{Quantum gates and unitary errors.--}}
%
%
As benchmarks for our investigation, we will consider one- and two-qubit gates. Then, we compare their capability to generate coherences dynamically, when perfectly implemented and when affected by a specific, yet experimentally relevant, source of imperfections. 

When considering the single-qubit operations, we refer to the unitary transformation
\begin{equation}
\label{Rop}
R(\theta)=e^{i\frac{\pi}{2}(I-{\bm n}\cdot{\bm \sigma})}=\cos(2\theta)\sigma_z+\sin(2\theta)\sigma_x \,,
\end{equation}
where ${\bm\sigma} \equiv (\sigma_x,\sigma_y,\sigma_z)$ the vector of Pauli operators $\sigma_k$ ($k=x,y,z$) and $I$ the $2\times2$ identity operator. Eq.~\eqref{Rop} embodies a rotation of an angle $\pi$ about the axis identified by the vector ${\bm n} \equiv (\sin(2\theta),0,\cos(2\theta))$. Such transformation is also used to construct the conditional two-qubit gate
\begin{equation}
G(\theta) \equiv \sigma^a_{+} \otimes I^b + \sigma^a_{-} \otimes R^b(\theta)
\end{equation}
with $\sigma^j_\pm=(\sigma^j_x\pm i\sigma^j_y)/2$ denoting the ladder operators and $j=a,b$ the label for the qubits being considered. The states of the computational basis for each qubit are $\{\ket{0}_j,\ket{1}_j\}$ where $\sigma^j_z\ket{k}_j=(-1)^k\ket{k}_j~(k=0,1)$. The conditional gate $G(\theta)$ applies a rotation of $\pi$ around axis ${\bm n}$ (or the identity matrix) to the state of qubit $b$ when the control qubit is in the state $\ket{1}_a$ (or in the state $\ket{0}_a)$, while the logical state of the control qubit is simultaneously flipped. 

The implementation of quantum gates is often affected by experimental imperfections that give rise to computational errors~\cite{KnillNature2005,KuengPRL2016,GeorgopoulosPRA2021}. In general, if we denote by $U=\{R(\theta),G(\theta)\}$ the perfect target gate we want to implement and ${\cal U}(\rho)=U\rho U^\dag$ the corresponding unitary map, its realization prone to errors will be denoted by $({\cal E}\circ{\cal U})(\rho)\equiv\mathcal{E}\left(\mathcal{U}(\rho)\right)$, where $\mathcal{E}(\cdot)$ is a completely positive trace preserving (CPTP) channel. The latter represents the error in the gate implementation. Among the most insidious and important errors are unitary ones, i.e., errors that do not affect the purity of the state. They result in 
\begin{equation}
({\cal E}\circ{\cal U})\rho = V\rho V^\dag,
\end{equation}
where $V$ is a unitary operation characterising the noisy channel. In our study, we will focus our attention on two experimentally relevant classes of unitary errors, namely \emph{rotation-angle errors} and \emph{rotation-axis errors}. The latter are described by
\begin{equation}
    V_{\rm axis}(\theta,\phi) = \sigma^a_{+} \otimes I^b +i \sigma^a_{-} \otimes \widetilde{R}^b_{\rm axis}(\theta,\phi)
\end{equation}
where we have introduced the rotation $\widetilde{R}_{\rm axis}(\theta,\phi) \equiv -i(\tilde{\bm n}\cdot{\bm \sigma})$ with $\tilde{\bm n} \equiv (\sin(2\theta)\cos(\phi),\sin(\phi),\cos(2\theta)\cos(\phi))$ denoting the rotation axis that {\it differs} from ${\bm n}$ by an angle $\phi$. Such error leaves the rotation angle unaffected. On the other hand, when considering rotation-angle errors, we look into 
\begin{equation}
    V_{\rm angle}(\theta,\varphi) = \sigma^a_{+} \otimes I^b + \sigma^a_{-} \otimes \widetilde{R}^b_{\rm angle}(\theta,\varphi)
\end{equation}
with
\begin{equation}
    \widetilde{R}_{\rm angle}(\theta,\varphi) \equiv
    i\cos(\alpha)I + \sin(\alpha)[\cos(2\theta)\sigma_z+\sin(2\theta)\sigma_x],  
\end{equation}
where $\alpha \equiv (\varphi+\pi)/2$. It is worth observing that these errors essentially affect only the target qubit $b$ and, as such, can be effectively considered as single-qubit errors.

\noindent
\textbf{\textit{Figures of merit.-}}
%
%
The quantification of the errors affecting a quantum gate is paramount to the achievement of fault-tolerant quantum computing. In this regard, without resorting to more expensive techniques as artificial intelligence ones~\cite{HarperNatPhys2020,MartinaQMI2022}, a typical way to estimate errors is through figures of merit such as the \emph{average gate fidelity}~\cite{NielsenPLA2002,VanicekPRE2006}
\begin{equation}
\label{avfid}
    \mathcal{F}\left(\mathcal{E}\circ\mathcal{U},\mathcal{U}\right) \equiv \mathbb{E}_{\ket{\psi}}{\rm Tr}
    \left[(\mathcal{E}\circ{\cal U})(\rho_{\ket{\psi}})\,{\cal U}(\rho_{\ket{\psi}}) \right],
\end{equation}
where the symbol $\mathbb{E}_{\ket{\psi}}$ stands for the ensemble average over all pure initial states $\ket{\psi}$ such that $\rho_{\ket{\psi}} \equiv |\psi\rangle\!\langle\psi|$. Such quantum states are drawn uniformly, according to the Haar measure, from the state space. 

While allowing for a coarse grained characterization of the quality of a quantum gate, Eq.~\eqref{avfid} requires the experimentally demanding tomographic reconstruction of the maps being involved~\cite{Chuang1997,Dariano2001,Mohseni2008}. Moreover, quantum states and processes that, according to fidelity-based figures or merit, are deemed to be close to each other might be endowed with significantly different physical properties~\cite{Mandarino2016}, thus weakening the foundations of any comparison based on quantities akin to Eq.~\eqref{avfid}. Finally, $\mathcal{F}\left(\mathcal{E}\circ\mathcal{U},\mathcal{U}\right)$ would not allow to easily single out the {\it quality} of the experimental process in regard to the generation of quantum coherence.

In order to bypass such bottlenecks, we put forward a figure of merit that more fits to the purpose at the core of our investigation. We thus consider the estimator of the mismatch between the quantum coherence generated by the gates affected by unitary errors and the noiseless ones respectively. Such estimator is defined as 
\begin{equation}
\label{eq:average_gate_coh_fid}
    \mathfrak{C}\left(\mathcal{E}\circ\mathcal{U},\mathcal{U}\right) \equiv
    \mathbb{E}_{\ket{\psi}}
    \Big[\left| C_{\ell_1}[ (\mathcal{E}\circ\mathcal{U}) (\rho_{\ket{\psi}})] - C_{\ell_1}[ \mathcal{U}(\rho_{\ket{\psi}})]\right|\Big],
\end{equation}
where $C_{\ell_1}[\rho] \equiv \sum_{n\neq k}\left|\rho_{nk}\right|$ denotes the $\ell_1$ measure of quantum coherence of the generic state $\rho$~\cite{BaumgratzPRL2014}. The average is again performed over all pure initial states in order to remove any dependence of the quantifier from the specific state that one may consider. We dub Eq.~\eqref{eq:average_gate_coh_fid} the \emph{average gate coherence fidelity}. It quantifies how much the coherence content of the quantum state after the application of the gate changes, on average, due to the presence of errors. 
While Eq.~\eqref{eq:average_gate_coh_fid} has a clear interpretation, determining  $C_{\ell_1}$ is in general not a trivial task. In fact, it would require either the tomographic reconstruction of the states -- and thus the evaluation of their degree of coherence -- or the use of entangled measurements on two copies of each state, in line with Ref.~\cite{Yuan2020}.

We now show that a different quantity, with a clear operational meaning and a fundamentally {\it local} nature, is closely connected with the average gate coherence fidelity for the gates and errors that we are considering, thus offering a route to its quantification. We focus our attention on the statistics of the energy fluctuations resulting from any physical mechanism of information processing, and thus also those at hand here. We look for the probability that, upon implementing a given dynamical process, the computational register under scrutiny is subjected to a change of its energy. We make use of the so-called end-point measurement (EPM) scheme~\cite{Ste_Ale_arXiv2021}, which has been recently formulated with the deliberate mandate of highlighting the role played, in such dynamical energy fluctuations, by quantum coherences. For a quantum system prepared in a state $\rho_0$ and subjected to a CPTP map $\mathcal{M}_t$ (here $t$ identifies the instant of time in the dynamics), the EPM's probability density function (PDF) for a given energy change is defined as 
\begin{equation}\label{eq:pdf}
    p_\text{EPM}\left(\Delta E_{k,\ell}\right) {=} p(E_{\rm in}^{k})\,p(E_{\rm fin}^{\ell}) 
    {=} {\rm Tr}[\Pi_{\rm in}^{k}\rho_0]{\rm Tr}[\Pi_{\rm fin}^{\ell}\mathcal{M}_{t_{\rm fin}}(\rho_0)],
\end{equation}
where $\Pi_{\rm in(fin)}^{j}$ are the projectors on the $j^{\rm th}$ initial (final) energy eigenstate of the system, and $\Delta E_{k,\ell} = E_{\rm fin}^{\ell} {-} E_{\rm in}^{k}$ is the corresponding energy change. $E_{\rm in}^{k}$ ($E_{\rm fin}^{\ell}$) denotes the eigenvalues of the initial (final) Hamiltonian. Then, we introduce the characteristic function of the EPM distribution: $\mathcal{G}(u) \equiv \langle e^{iu\Delta E}\rangle_{\rm EPM} = {\rm Tr}[e^{-iu H_{t_{\rm in}}}\rho_0]{\rm Tr}[e^{iu H_{t_{\rm fin}}}\mathcal{M}_{t_{\rm fin}}(\rho_0)]$ with $H_t$ the (time-dependent) Hamiltonian of the system and $t_{{\rm in}({\rm fin})}$ the initial (final) time of the process. The EPM characteristic function can be cast as $\mathcal{G}(u;{\cal M}_t)=\sum_{{\cal Q}={\cal P},\chi}\mathcal{G}_{\mathcal{Q}}(u;{\cal M}_t)$ with
\begin{equation}\label{eq:charact_function}
\begin{aligned}
    \mathcal{G}_{{\cal Q}}(u;{\cal M}_t)&= {\rm Tr}[e^{-iu H_{t_{\rm in}}}\rho_0]\,{\rm Tr}[e^{iu H_{t_{\rm fin}}}\mathcal{M}_{t_{\rm fin}}({\cal Q})]
\end{aligned}
\end{equation}
and $u\in\mathbb{C}$. In Eq.~(\ref{eq:charact_function}) we have decomposed the initial state as $\rho_0={\cal P}+\chi$ with the diagonal part $\mathcal{P}$ (in the basis of the initial Hamiltonian $H_{t_{\rm in}}$) and the traceless component $\chi$ that encodes the quantum coherence in the energy basis. We can thus isolate the contribution $\mathcal{G}_{\chi}(u;{\cal M}_t)$ of the characteristic function stemming from the initial quantum coherence.

\begin{figure*}[t!]
\centering
{(a)}\hskip4cm{(b)}\hskip4cm{(c)}\hskip4cm{(d)}\\
\includegraphics[width=2\columnwidth]{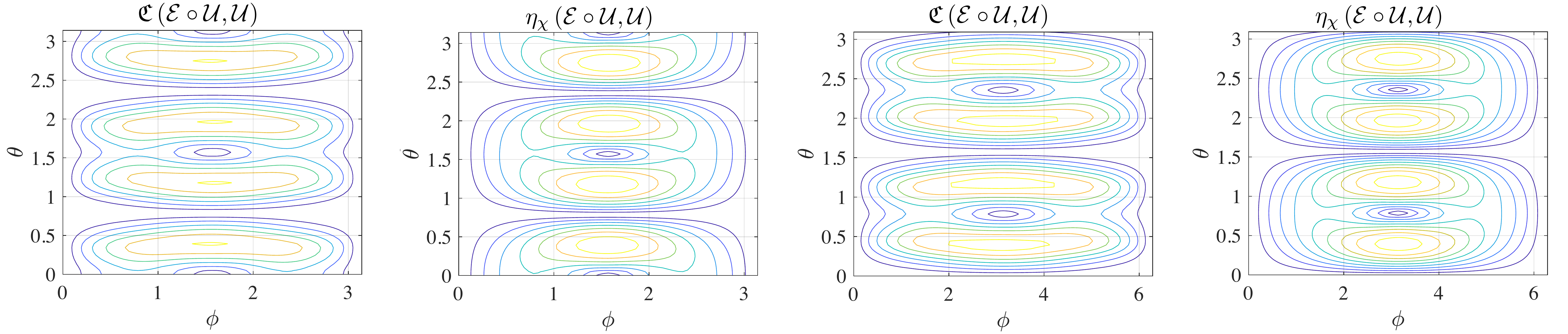}
\caption{{\bf Two-qubit gate, rotation angle and axis errors}. We show $\mathfrak{C}\left(\mathcal{E}\circ\mathcal{U},\mathcal{U}\right)$ and $\eta_{\chi}({\cal E}\circ{\cal U},\mathcal{U})$, both averaged over $5\times10^3$ initial random pure states, against both the perfect gate angle $\theta\in[0,\pi]$ and the error parameter $\phi$. We take $\phi\in[0,\pi]$ in case of rotation-axis error [panels {(a)} and {(b)}] and $\phi\in[0,2\pi]$ for the rotation-angle error [panels {(c)} and {(d)}].}
\label{fig:2qubit_unitary_errors}
\end{figure*} 

In the context of the problem addressed by our study, we consider the statistics of the local energy originated by the Hamiltonian $H=\sigma^a_z\otimes{I}^b+{I}^a\otimes\sigma^b_z$. This just entails local measurements in the computational basis~\cite{CimininpjQI2020,Ste_Ale_arXiv2021}, embodying a remarkable simplification of the estimation process. Then, we consider the following figure of merit:
\begin{equation}
\label{figmer}
    \eta_{\chi}\left(\mathcal{E}\circ\mathcal{U},\mathcal{U}\right) \equiv \mathbb{E}_{\ket{\psi}}\Big[\big|\mathcal{G}_{\chi}(i;\mathcal{E}\circ\mathcal{U})-\mathcal{G}_{\chi}(i;\mathcal{U})\big|\Big].
\end{equation}
Eq.~\eqref{figmer} is built upon the difference between the coherence-dependent components of the EPM characteristic function that result from the local-energy PDF corresponding to the ideal (target) gate and its error-affected version.

An extensive numerical investigation of the considered gates and errors demonstrates a remarkable alignment of the behaviour of $\eta_{\chi}({\cal E}\circ{\cal U},{\cal U})$ and $\mathfrak{C}({\cal E}\circ{\cal U},{\cal U})$, as shown in Fig.~\ref{fig:2qubit_unitary_errors}. As a result, when averaging over the pure input states, the difference in the statistics of the local EPM energy changes that accounts for the contribution from $\chi$ reproduces the behaviour of the average gate coherence fidelity for two-qubit controlled gates affected by either a rotation-angle or rotation-axis error. However, this is not in general recovered under generic maps affecting the ideal gates. In the Supplementary Material accompanying the manuscript, we address analytically the figures of merit at the core of our study.

\noindent
\textbf{\textit{Experimental results.-}}
%
\begin{figure}[t!]
    \includegraphics[width=0.95\columnwidth]{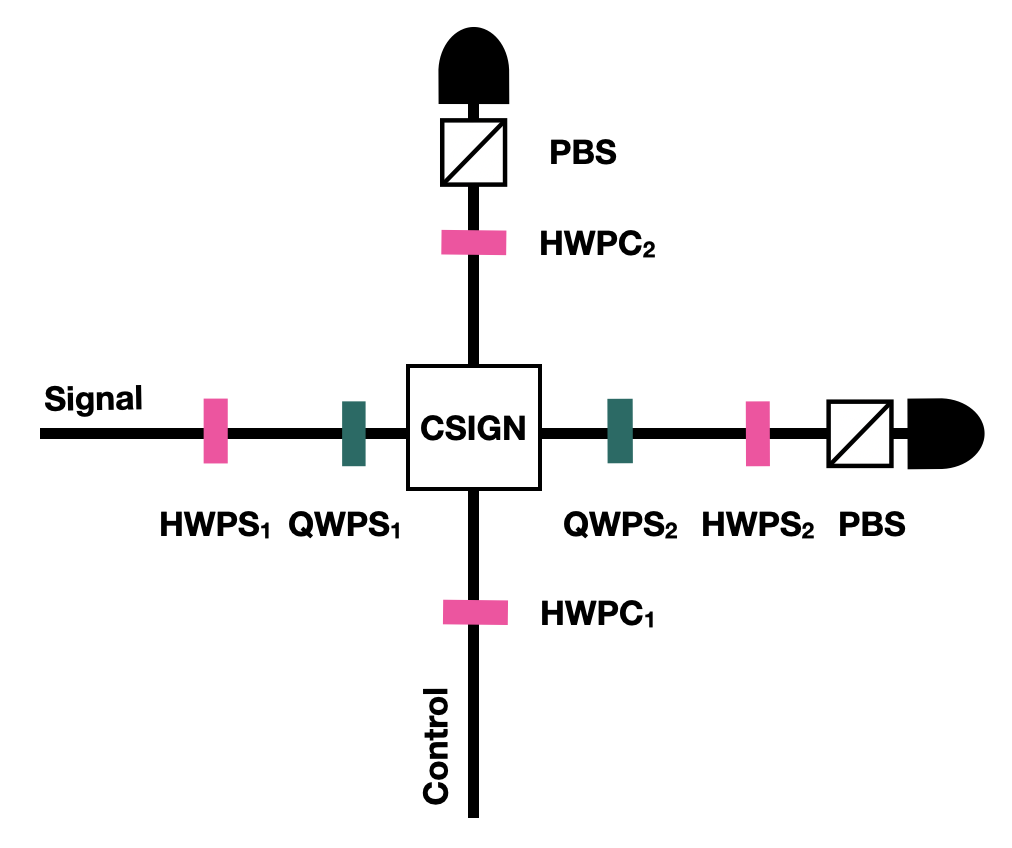}
    \caption{The two photons are generated through Type I SPDC on a 3 mm BBO crystal pumped with a 405 nm CW laser. The photons are filtered with a FWHM 7.5 nm interference filter and sent to the setup depicted in the figure through single mode fibres. The photons are prepared in the $|0\rangle,|1\rangle,$ or $|+\rangle$ states using ${\rm HWPC}_1$ and ${\rm HWPS}_1$ respectively. {The waveplates on the signal arm are used  to implement unitary errors in a controllable way, and the preparation and measurement settings are included in the overall transformation; this is not necessary on the control arm, for which the plates control the preparation and the measurement directly}. The C-SIGN gate is constituted by a main PPBS transmitting 1/3 and reflecting 2/3 of the vertically polarized light. There are also two additional PPBS -- one per arm -- that are rotated by 90 degrees, thus operating on the H polarization. The additional PPBS allow to compensate for the intensity unbalance generated by the main PPBS.}
    \label{fig:setup}
\end{figure}
\begin{figure}[t!]
\centering
\includegraphics[width=0.95\columnwidth]{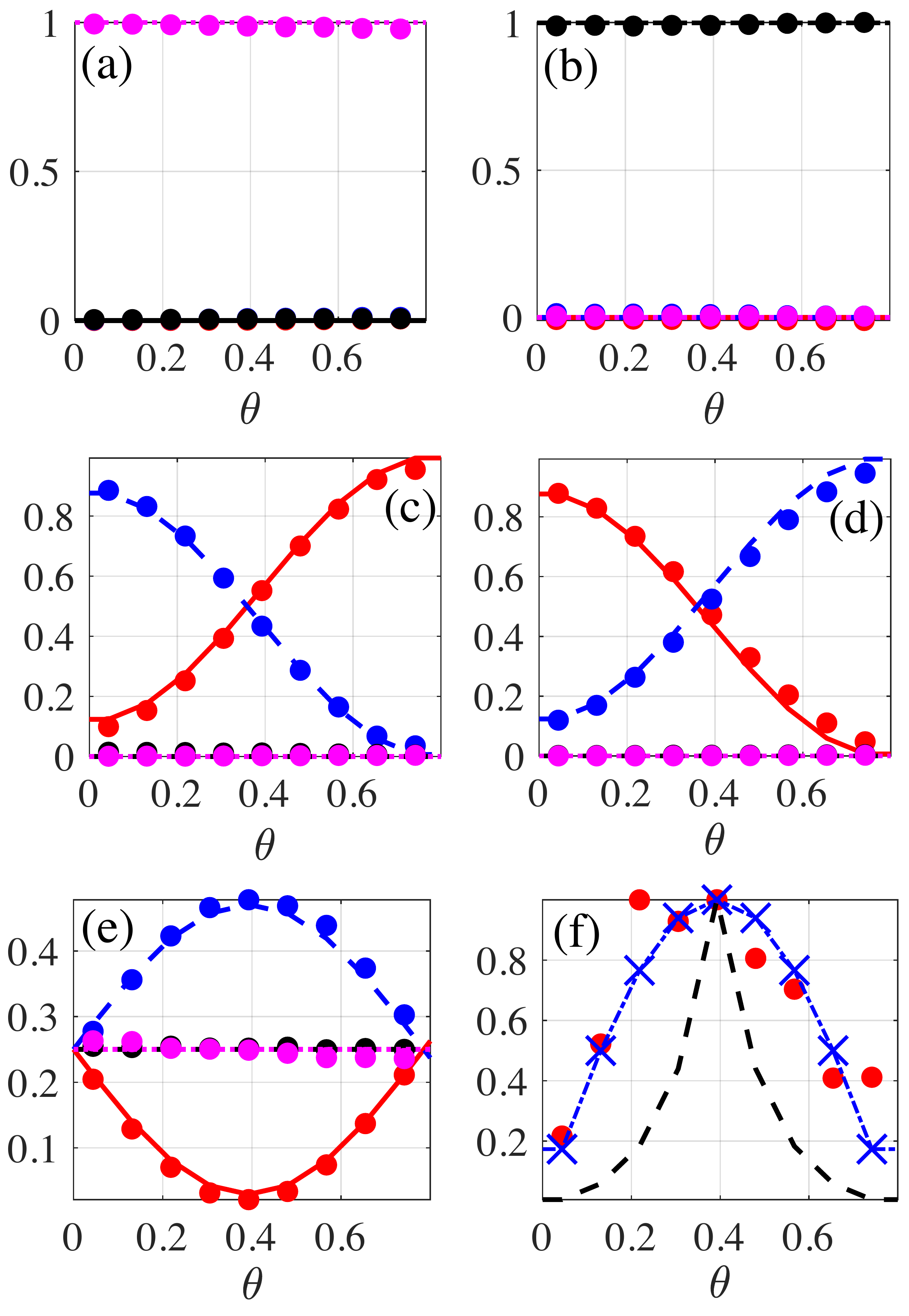}
\caption{
In the panels (a)-(e), the dots represent the experimental data, while the solid lines denote the theoretical predictions. Specifically, (a) Red: $p(00|00)$, Blue: $p(01|00)$, Black: $p(10|00)$, Magenta: $p(11|00)$; (b) Red: $p(00|01)$, Blue: $p(01|01)$, Black: $p(10|01)$, Magenta: $p(11|01)$; (c) Red: $p(00|10)$, Blue: $p(01|10)$, Black: $p(10|10)$, Magenta: $p(11|10)$; (d) Red: $p(00|11)$, Blue: $p(01|11)$, Black: $p(10|11)$, Magenta: $p(11|11)$; (e) Red: $p(00|++)$, Blue: $p(01|++)$, Black: $p(10|++)$, Magenta: $p(11|++)$. (f) Red dots: inference of $1.6153\big|\mathcal{G}_{\chi}(i;\mathcal{E}\mathcal{U})-\mathcal{G}_{\chi}(i;\mathcal{U})\big|$ with experimental data, Blue x-mark dash-dotted line: theoretical prediction of $2.24774\big|\mathcal{G}_{\chi}(i;\mathcal{E}\mathcal{U})-\mathcal{G}_{\chi}(i;\mathcal{U})\big|$, Black dashed line: $3.7404\big| C_{\ell_1}( \mathcal{E}\mathcal{U} ) - C_{\ell_1}( \mathcal{U} )\big|$. All the curves in the panels are plotted as a function of $\theta\in[0,\pi/4]$ rad and $\phi = \pi/9$ rad, and by initializing the implemented quantum gate in the single initial state $\ket{++}$.}
\label{fig:exp_results}
\end{figure}
We will now test our numerical prediction using a quantum optics implementation of the controlled two-qubits gate affected by rotation axis errors.

We encode the logical states $|0\rangle_j$ and $|1\rangle_j$ in the horizontal and vertical polarization states $|H\rangle_j$ and $|V\rangle_j$ of a photon (here, $j=a,b$). The realization of the two-qubit gate is based on the design of a controlled-sign gate illustrated in Fig.~\ref{fig:setup}. In order to implement the map associated with rotation-axis errors $V_{\rm axis}(\theta,\phi)$, we perform rotations on the polarization of the signal photon using a set of half (HWP) and quarter (QWP) wave plates before and after the interaction between the signal and control photons. For this purpose, we set the angle of both HWPs to $\alpha_{{\rm HWPS}_{1,2}} = \theta/2 + \phi/4$ and the angle of the QWPs to $\beta_{{\rm QWPS}_1} = \phi/2 + \pi/2$ and $\beta_{{\rm QWPS}_2} = \phi/2$ respectively. 

Remarkably, the expected behaviour of the gate allows to capture the essential features of the EPM-based diagnostics by focussing on a reduced set of input states:
in different experiments, we initialize the two-qubit gate in $\ket{++} \equiv (|0\rangle + |1\rangle)\otimes (|0\rangle + |1\rangle)/2$ as well as in $|00\rangle,|01\rangle,|10\rangle,|11\rangle$. These choices provide enough settings to compare coherent and incoherent cases with figures close to the averaged one. As an illustrative example, we fix the value of $\phi=20^{\circ}$ by letting vary $\theta$ between $0^{\circ}$ and $45^{\circ}$. Finally, in output of the two-qubit gate, we collect sets of measurement outcomes by projecting on the basis $\{|00\rangle,|01\rangle,|10\rangle,|11\rangle\}$.  

In Fig.~\ref{fig:exp_results} we compare our theoretical predictions with the experimental data. In the previous section, we have discussed how the figure of merit $\eta_{\chi}$ bears close resemblances to the average gate coherence fidelity $\mathfrak{C}$ upon the average over all initial pure states. The experimental determination of the average estimator lays beyond the scope of our current work (cf. the Supplementary Material for an additional analysis). Nonetheless, with our restricted set of initialization states we can faithfully reconstruct the kernel $|\mathcal{G}_{\chi}(i;\mathcal{E}\mathcal{U})-\mathcal{G}_{\chi}(i;\mathcal{U})|$ of our estimator for the initial state $\rho_0 = \ket{++}\!\bra{++}$ and compare it with the corresponding gate coherence fidelity $|C_{\ell_1}(\mathcal{E}\mathcal{U}) - C_{\ell_1}(\mathcal{U})|$ for the same initial state [cf. Fig.~\ref{fig:exp_results}(f)]. 

These measurements support our previous observation on how, beyond convenience,  the quantities estimated in this way have a qualitatively similar behavior to the averaged figures of merit $\eta_{\chi}$ and $\mathfrak{C}$. The observed discrepancies are not due to the lack of averaging, but flag genuine non-idealities of the gate adding up to the unitary error. Specifically, imperfect non-classical visibility for the vertical polarizations, as well as residual unwanted interference for the horizontal polarizations are responsible for these extra faults~\cite{CimininpjQI2020}.

Note that, as the probabilities to measure the $\ell^{\rm th}$ local energy of the gate  at time $t_{\rm fin}$ is also equal to
\begin{equation}
\begin{aligned}
    p(E_{\rm fin}^{\ell};{\cal Q}=\chi)&={\rm Tr}[|kq\rangle\!\langle kq|\mathcal{M}_{t_{\rm fin}}(\chi)] \\
    &=p(kq|++) - \displaystyle{\frac{1}{4}\sum_{n,m=0}^{1}p(kq|nm)}
    \end{aligned}
\end{equation}
with $k,q=0,1$, the contribution $\mathcal{G}_{\chi}$ of the EPM characteristic function is reconstructed experimentally from measuring the conditional probabilities $p(00|kq)$, $p(01|kq)$, $p(10|kq)$, $p(11|kq)$, $p(kq|++)$ (panels (a)-(e) in Fig.~\ref{fig:exp_results}). Conversely, the $\ell_{1}$ measure of quantum coherence $C_{\ell_{1}}$ are obtained by numerical simulations as a function of $\theta\in[0,\pi/4]$ rad and $\phi=\pi/9$ rad.\\

\noindent
\textbf{\textit{Conclusions.-}}
%
%
In this paper we introduce and discuss a tool for the diagnostics of one- and two-qubit gates subjected to unitary errors that are typically cumbersome~\cite{KnillNature2005,KuengPRL2016}. Specifically, we point out how an estimator ($\eta_{\chi}$) based on the recently introduced end-point measurement (EPM) scheme~\cite{Ste_Ale_arXiv2021}, compares with the average gate coherence fidelity $\mathfrak{C}$. The determination of our estimator requires only local energy measurements of the qubits output states, and we show that it qualitatively reproduces the average gate coherence fidelity. We also compare the estimator and the gate coherence fidelity on a single initial state by using an all-optical set-up.

Our results employ thermodynamics tools, i.e., the characteristic function of the EPM energy change statistics, to investigate the faithfulness of quantum logic gates. This is indeed a growing research field that is attracting the attention of the wider quantum community. For our case, beyond requiring only local energy measurements, the EPM approach provides a complementary way to perform diagnostics of quantum-gate coherences without resorting to tomographic procedures.

It would  be interesting to investigate the general properties of $\eta_{\chi}$ as estimator of quantum coherence for quantum technology applications that make use of it as a resource, from quantum communication to quantum batteries~\cite{batteries}, quantum transport~\cite{Benenti}, and clocks~\cite{woods2016}.\\
\noindent
\textbf{\textit{Acknowledgments.-}}
We acknowledge support from the European Union's Horizon 2020 FET-Open projects TEQ (Grant Agreement No.\,766900), and STORMYTUNE (Grant Agreement No.\,899587), the Leverhulme Trust Research Project Grant UltraQuTe (grant RGP-2018-266), the Royal Society Wolfson Fellowship (RSWF/R3/183013), the UK EPSRC (grant EP/T028424/1), the Department for the Economy Northern Ireland under the US-Ireland R\&D Partnership Programme, the Deutsche Forschungsgemeinschaft (DFG, German Research Foundation) project number BR 5221/4-1, the Blanceflor Foundation for financial support through the project ``The theRmodynamics behInd thE meaSuremenT postulate of quantum mEchanics (TRIESTE)''.


\newpage
\widetext
\quad
\begin{center}
\textbf{\large Supplemental Material: Diagnostics of quantum-gate coherences via end-point-measurement statistics}
\end{center}
\setcounter{equation}{0}

\setcounter{figure}{0}
\setcounter{table}{0}
\setcounter{page}{1}
\makeatletter
\renewcommand{\theequation}{S\arabic{equation}}
\renewcommand{\thefigure}{S\arabic{figure}}
\renewcommand{\bibnumfmt}[1]{[S#1]}
\renewcommand{\citenumfont}[1]{S#1}

\section*{Figures of merit: Formal analysis}

In this section, we take a better look at the quantities of interest in the main text of the paper. Let us thus take into account the expressions of the considered figures of merit before performing the averaging over the initial pure states $\rho_0 = \sum_{n,m}(\rho_0)_{nm}|n\rangle\!\langle m|$ with $(\rho_0)_{nm} \equiv a_n^{*}a_m$ and $\ket{\psi_0} \equiv \sum_i a_i\ket{i}$. 

\subsection{1. Expressions of $\eta_{\rm TPM}$ and $\eta_{\mathcal{P}}$}

\begin{itemize}
    \item 
    If one made use of the two-point measurement (TPM) scheme, then a possible figure of merit for the diagnostic of a quantum gate would be
\begin{align}
    \eta_{\rm TPM}&=\mathbb{E}_{\ket{\psi_0}}\Big[\big|{\rm Tr}\left(e^{-H_{\rm fin}}\left(\mathcal{E}(U e^{H_{\rm in}}\mathcal{P}U^{\dag})-Ue^{H_{\rm in}}\mathcal{P}U^{\dagger}\right)\right)\big|\Big] = \mathbb{E}_{\ket{\psi_0}}\left[\aleph_{\rm TPM}\right].
\end{align}
Using the computational basis (over which the observable $H=H_{\rm in}=H_{\rm fin}$ is diagonal such that $A_k \equiv \langle k|H|k\rangle$), one then gets  
\begin{align}
    \aleph_{\rm TPM}=\left|\sum_{n,\alpha,m} e^{-H_n}e^{H_m}(\rho_{0})_{mm}\left(\bra{n}K_\alpha U\ket{m}\bra{m}U^\dag K_\alpha^\dag\ket{n}-\bra{n} U\ket{m}\bra{m}U^\dag\ket{n}\right)\right|,
\end{align}
where $(\rho_{0})_{mm} = |\bra{m}\psi_0\rangle|^2$, and we have expressed the noise channel affecting the quantum gate in term of the corresponding Kraus representation: $\mathcal{E}(\cdot)=\sum_\alpha K_\alpha (\cdot) K_\alpha^\dag$. In the case of a unitary error, just a single Kraus operator is different from zero.
\item 
In the case of $\mathcal{G}_{\mathcal{P}}$ one has
\begin{align}
    \eta_{\mathcal{P}}=\mathbb{E}_{\ket{\psi_0}}\Big[\big|\langle\psi_0|e^{H_{t_{\rm in}}}|\psi_0\rangle{\rm Tr}\left(e^{-H_{t_{\rm fin}}}\left(\mathcal{E}(U\mathcal{P}U^{\dag})-U\mathcal{P}U^{\dagger}\right)\right)\big|\Big]=\mathbb{E}_{\ket{\psi_0}}\left[\aleph_{\mathcal{P}}\right].
\end{align}
In this case, still using the computational basis (over which the observables $H=H_{\rm in}=H_{\rm fin}$ and $\mathcal{P}$ are diagonal), we obtain  
\begin{align}
    \aleph_{\mathcal{P}}&=\langle e^{H}\rangle\left|\sum_{n,\alpha} e^{-H_n}\Big(\bra{n}K_\alpha U\mathcal{P}U^\dag K_\alpha^\dag\ket{n}-\bra{n} U\mathcal{P}U^\dag\ket{n}\Big)\right|\nonumber \\
    &=\langle e^{H}\rangle\left|\sum_{n,\alpha,m} e^{-H_n}(\rho_0)_{mm}\Big(\bra{n}K_\alpha U\ket{m}\bra{m}U^\dag K_\alpha^\dag\ket{n}-\bra{n} U\ket{m}\bra{m}U^\dag\ket{n}\Big)\right|.
\end{align}
\end{itemize}
From these expressions we see immediately that $\aleph_{\rm TPM}$ and $\aleph_{\mathcal{P}}$ essentially encode the same information. In fact, they take into account the differences between the same diagonal and off-diagonal elements of the quantum states after the action of the perfect and imperfect/noisy gates (with different weights though).

\subsection{2. Average gate fidelity}

The average gate fidelity is defined as
\begin{equation}
    \mathcal{F}\left(\mathcal{E}\circ \mathcal{U},\mathcal{U}\right) \equiv \mathbb{E}_{\ket{\psi_0}}\left[ \bra{\psi}U^\dag\mathcal{E}(U|\psi\rangle\!\langle\psi|U^{\dag})U\ket{\psi} \right]=\mathbb{E}_{\ket{\psi_0}}\left[\mathfrak{F}\right],
\end{equation}
where
\begin{align}
    \mathfrak{F}&=\sum_{n_1,n_2,m_1,m_2,\ell_1,\ell_2,\alpha}(\rho_0)_{n_1n_2}(\rho_{0})_{\ell_1\ell_2}\bra{n_1}U^\dag\ket{m_1}\bra{m_1}K_\alpha U\ket{\ell_1}\bra{\ell_2}U^\dag K_\alpha^\dag\ket{m_2}\bra{m_2}U\ket{n_2}.
\end{align}

\subsection{3. Expressions of $\eta_{\rm EPM}$ and $\eta_{\chi}$}

\begin{itemize}
    \item 
    The figure of merit, using the (full) EPM energy statistics, for the diagnostics of a quantum gate is
\begin{align}
    \eta_{\rm EPM}=\mathbb{E}_{\ket{\psi_0}}\Big[\big|\langle\psi_0|e^{H_{t_{\rm in}}}|\psi_0\rangle{\rm Tr}\left(e^{-H_{t_{\rm fin}}}\left(\mathcal{E}(U|\psi_0\rangle\!\langle\psi_0|U^{\dag})-U|\psi_0\rangle\!\langle\psi_0|U^{\dagger}\right)\right)\big|\Big]=\mathbb{E}_{\ket{\psi_0}}\left[\aleph_{\rm EPM}\right].
\end{align}
Using the computational basis (over which the observable $H=H_{\rm in}=H_{\rm fin}$ is diagonal) as before, $\aleph_{\rm EPM}$ can be written as   
\begin{align}
    \aleph_{\rm EPM}&=\langle e^{H}\rangle\left|\sum_{n,\alpha} e^{-H_n}\Big(\bra{n}K_\alpha U\ket{\psi_0}\bra{\psi_0}U^\dag K_\alpha^\dag\ket{n}-\bra{n} U\ket{\psi_0}\bra{\psi_0}U^\dag\ket{n}\Big)\right|\nonumber \\
    &=\langle e^{H}\rangle\left|\sum_{n,\alpha,m_1,m_2} e^{-H_n}(\rho_0)_{m_1 m_2}\Big(\bra{n}K_\alpha U\ket{m_1}\bra{m_2}U^\dag K_\alpha^\dag\ket{n}-\bra{n} U\ket{m_1}\bra{m_2}U^\dag\ket{n}\Big)\right|
\end{align}
with $(\rho_{0})_{m_1 m_2} = \bra{m_1}\psi_0\rangle\langle\psi_0\ket{m_2}$ and $\langle e^{H}\rangle = \sum_m |\bra{m}\psi_0\rangle|^2 e^{H_m}$.
\item 
For the coherence part of $\eta_{\rm EPM}$, instead, one has 
\begin{align}
    \eta_{\chi}=\mathbb{E}_{\ket{\psi_0}}\Big[\big|\langle\psi_0|e^{H_{t_{\rm in}}}|\psi_0\rangle{\rm Tr}\left(e^{-H_{t_{\rm fin}}}\left(\mathcal{E}(U\chi U^{\dag})-U\chi U^{\dag}\right)\right)\big|\Big]=\mathbb{E}_{\ket{\psi_0}}\left[\aleph_{\chi}\right],
\end{align}
where
\begin{align}
    \aleph_{\chi}&=\langle e^{H}\rangle\left|\sum_{n,\alpha} e^{-H_n}\Big(\bra{n}K_\alpha U\chi U^\dag K_\alpha^\dag\ket{n}-\bra{n} U\chi U^\dag\ket{n}\Big)\right| \nonumber \\
    &=\langle e^{H}\rangle\left|\sum_{n,\alpha,m_1\neq m_2} e^{-H_n}(\rho_0)_{m_1 m_2}\Big(\bra{n}K_\alpha U\ket{m_1}\bra{m_2}U^\dag K_\alpha^\dag\ket{n}-\bra{n} U\ket{m_1}\bra{m_2}U^\dag\ket{n}\Big)\right|
\end{align}
obtained by using the computational basis over which the observable $H=H_{\rm in}=H_{\rm fin}$ is diagonal.
\end{itemize}

\subsection{4. Average gate coherence fidelity}

The average gate coherence fidelity is formally expressed by
\begin{equation}
    \mathfrak{C}\left(\mathcal{E}\circ\mathcal{U},\mathcal{U}\right) \equiv
    \mathbb{E}_{\ket{\psi_0}}\left[\left| C_{\ell_1}( \mathcal{E}\circ\mathcal{U} ) - C_{\ell_1}( \mathcal{U} )\right|\right]= \mathbb{E}_{\ket{\psi_0}}\left[\mathcal{C}\right],
\end{equation}
where
\begin{equation}
    \mathcal{C}=\left|\sum_{n\neq k,\alpha,m_1,m_2}\Big(\big|(\rho_0)_{m_1 m_2}\bra{n}K_\alpha U\ket{m_1}\bra{m_2}U^\dag K_\alpha^\dag\ket{k}\big| - \big|(\rho_0)_{m_1 m_2}\bra{n}U\ket{m_1}\bra{m_2}U^\dag\ket{k}\big|\Big)\right|.
\end{equation}
Accordingly, in general, the similarities between  $\eta_{\chi}$ and $\mathfrak{C}$ described in the main text stem from the averaging over the initial pure states $\rho_0$, though the analytical expressions for $\eta_{\chi}$ and $\mathfrak{C}$ (after the averaging) are not at our disposal. However, remarkable similarities in the behaviour of the two figures of merit can be observed even at the single-state level. Such similarities have been used in the main text, for example, when comparing theoretical predictions with experimental data obtained by initializing the two-qubit gate in the $\ket{++}$ state.

\section{Ways to determine experimentally the EPM-based figure of merit}

Let us focus on the case of two-qubit quantum gates and consider only the EPM final probabilities to measure the corresponding local energy terms
\begin{equation}
    p_{E}^{(f)}={\rm Tr}\left[\Pi_E^{(f)}U|\psi_0\rangle\!\langle\psi_0|U^\dag\right]=\bra{\psi_0}U^\dag\Pi_E^{(f)}U\ket{\psi_0}.
\end{equation}
We also have
\begin{equation}
    \mathcal{G}^{(f)}(\ket{\psi_0})=\bra{\psi_0}U^\dag e^{-H_{f}}U\ket{\psi_0}=\sum_\alpha e^{-E_\alpha}\bra{\psi_0}U^\dag\Pi_\alpha^{(f)}U\ket{\psi_0}
\end{equation}
which, for noisy unitary gates $U$ affected by coherent errors, is what should be determined experimentally.
In this regard, let us expand the initial state in the computational basis $\{\ket{i}\}=\{\ket{00},\ket{01},\ket{10},\ket{11}\}$. Then 
\begin{equation}
    \mathcal{G}^{(f)}(\ket{\psi_0})=\sum_\alpha e^{-E_\alpha}\sum_{ij}a_ia_j^*\bra{j}U^\dag\Pi_\alpha^{(f)}U\ket{i}.
\end{equation}
Thus, we need all the different quantities $\bra{j}U^\dag\Pi_\alpha^{(f)}U\ket{i}$ if we want to be able to determine (by post-processing) the averaged estimator that we are interested in. These  are $4\times (4+6\times 2) = 64$ terms, i.e., for each of the four energy eigenvalues $E_{\alpha}$ of the local, final Hamiltonian we have
\begin{itemize}
    \item Four real quantities $\bra{i}U^\dag\Pi_\alpha^{(f)}U\ket{i}$
    \item Six complex quantities $\bra{j}U^\dag\Pi_\alpha^{(f)}U\ket{i}$ with $i\neq j$.
\end{itemize}

Now, by initializing the state in $\ket{\psi_0}=\ket{i}$ we can recover the four real quantities which are just the final EPM probabilities. In order to recover the others we need other initial states, which we are going to discuss below.

\subsection{1. Straightforward initialization}

A straightforward way is to initialize the two-qubit system in the 12 states $\ket{\psi_0}=(\ket{i}+\ket{j\neq i})/\sqrt{2}$ and $\ket{\psi_0}=(i\ket{i}+\ket{j\neq i})/\sqrt{2}$. In fact, by doing so the final EPM probabilities take the form
\begin{align}
    & p_E^{(f)}=\frac{1}{2}\bra{i}U^\dag\Pi_E^{(f)}U\ket{i}+\frac{1}{2}\bra{j}U^\dag\Pi_E^{(f)}U\ket{j}+{\rm Re}[\bra{i}U^\dag\Pi_E^{(f)}U\ket{j}]\\
    & p_E^{(f)}=\frac{1}{2}\bra{i}U^\dag\Pi_E^{(f)}U\ket{i}+\frac{1}{2}\bra{j}U^\dag\Pi_E^{(f)}U\ket{j}+{\rm Im}[\bra{i}U^\dag\Pi_E^{(f)}U\ket{j}].
\end{align}
This means that with a total of $4\times 4+4\times(6\times 2)=64$ projective local measurements we can get all the terms we need.
However, note that the additional 12 states needed here are, in general, entangled, which could be a complication for an optical set-up like the one considered in the main text.

\subsection{2. Working with separable input states}

If instead we want to use separable states then we could do as follow.
Let us start from the 4 states $\ket{\psi_0}=(\ket{aa}+\ket{ab})/\sqrt{2},\,(\ket{aa}+\ket{ba})/\sqrt{2}$ with $a,b\in \{0,1\}$ and $a\neq b$ and, correspondingly, the 4 states $\ket{\psi_0}=(\ket{aa}-i\ket{ab})/\sqrt{2},\,(\ket{aa}-i\ket{ba})/\sqrt{2}$ (for a total of 8 states providing us 4 complex quantities among the 6 $\bra{j}U^\dag\Pi_\alpha^{(f)}U\ket{i}$ with $i\neq j$). With these initialization, the computation of the EPM final probabilities again gives us, apart from diagonal terms, the real and imaginary parts of 4 of the terms $\bra{i}U^\dag\Pi_E^{(f)}U\ket{j}$. Then, one can consider starting from the $\ket{++}$ state which would give, apart from the diagonal terms and the terms that we can extract from the previous states, the term 
\begin{equation}\label{SM:eq_sum_real_terms}
    \frac{1}{2}\left({\rm Re}\bra{00}U^\dag\Pi_E^{(f)}U\ket{11}+{\rm Re}\bra{10}U^\dag\Pi_E^{(f)}U\ket{01}\right).
\end{equation}
Now, to conclude, one could start from an entangled state (like $\frac{1}{2}(\ket{00}-\ket{11}+\ket{10}+\ket{01})$ or $\frac{1}{2}(-\ket{00}+\ket{11}+\ket{10}+\ket{01})$) to also get the difference of the real parts in (\ref{SM:eq_sum_real_terms}). This whole process would require $8+4=12$ initial states as before. 

If we want to get rid also of the single Bell state that we need for the difference of the real parts, we could, e.g., start from 
\begin{equation}
\frac{1}{2}(i\,e^{i\theta}\ket{00}+i\ket{01}+e^{i\theta}\ket{10}+\ket{11})=\frac{1}{2}(i\ket{0}+\ket{1})\otimes (e^{i\theta}\ket{0}+\ket{1}),
\end{equation}
which is separable, and gives the difference of the real parts in (\ref{SM:eq_sum_real_terms}) for $\theta=\pi/2$. 

\end{document}